# Interleukin-2 receptor antagonists for pediatric liver transplant recipients: A systematic review and meta-analysis of controlled studies

Nicola D. Crins[1,2], Christian Röver[1], Armin D. Goralczyk[3], Tim Friede[1]

[1]Department of Medical Statistics, University Medical Center Göttingen, Göttingen, Germany
[2]Department of Internal Medicine, Clinical Center Wolfenbüttel, Wolfenbüttel, Germany
[3]Department of Internal Medicine, Clinical Center Herzberg, Herzberg, Germany

**Abstract**

Interleukin-2 receptor antagonists (IL-2RA) are frequently used as induction therapy in liver transplant recipients to decrease the risk of acute rejection while allowing the reduction of concomitant immunosuppression. The exact association with the use of IL-2RA however is uncertain. We performed a systematic literature search for relevant studies. Random effects models were used to assess the incidence of acute rejection, steroid-resistant rejection, graft loss, patient death, and adverse drug reaction, with or without IL-2RA. Six studies (2 randomized and 4 nonrandomized) met the elegibility criteria. Acute rejection at 6 months or later favored the use of IL-2RA significantly (relative risk (RR) 0.38; 95% confidence interval (CI) 0.22-0.66, p = 0.0005). Although not statistically significant, IL-2RA showed a substantial reduction of the risk of steroid-resistant rejection (RR 0.32; CI 0.19-1.03, p = 0.0594). Graft loss and patient death showed a reductive tendency through the use of IL-2RA. The use of IL-2RA is safe and is associated with a statistically significantly lower incidence of acute rejection after transplantation and substantial reduction of steroid-resistant rejection, graft loss and patient death.

## Introduction

The ultimate purpose of pediatric liver transplantation is to grant an expectance of life of several decades. Immunosuppression should be tailored to ensure best management of both short and long term complications. Currently the immunosppressive strategey is not standardized among different transplantation centers mainly because very few comparative studies with adequate number of patients and duration of follow-up are published (1-3). In particular, only a few published controlled clinical studies report on the use of a relatively new immunosuppressive agent called basiliximab (Simulect®) or daclizumab (Zenapax®). These are monoclonal antibodies targeting the interleukin-2 (IL-2) receptor (IL-2R). Initially they were approved for therapy of patients after renal transplantation. The two IL-2R antibodies (IL-2RA) daclizumab and basiliximab were commercially available, but daclizumab has recently been withdrawn from the market for commercial reasons. Now basiliximab is regulary used in adult as well as in pediatric liver transplant both in Europe (4) and in the US (5).

The aim of induction therapy with IL-2RA is mainly to decrease the risk of acute rejection. Acute rejection should be prevented because a graft is damaged with each rejection and loses part of its function. There are histopathologic features of acute and chronic rejection proved by core needle biopsy (6). Avoiding acute rejection (AR) or steroid-resistant (acute) rejection (SRAR) is improves the long time function of a liver graft. For children a good functioning and long living liver graft is particularly important due to the high expected lifespan. Acute rejection is a strong risk factor for chronic rejection in kidney transplant patients (7). Acute rejection after liver transplantation can progress to chronic injury, it shows a prolonged liver dysfunction, and all lead on to chronic rejection (6). It could be a similar strong risk factor for liver recipients, but the development is not well understood by now (8). Chronic rejection is also described as a potentially reversible process with a dynamic nature (8). There is no evidence based connection between acute rejection and long-term outcome in liver transplantation. FDA regulated clinical research focuses on acute rejection as primary endpoint, leading to less information about long-term data and other endpoints. The idea of using IL-2RA is an exchange of immunosuppressive drugs without increasing the risk of graft damage to reduce long lasting effects of common immunosuppressive substances such as steroids and calcineurin inhibitor (CNI). Common side effects of steroids include infections, arterial hypertension, glucose intolerance, hyperlipidemia and osteoporosis (9). There are some side effects which differ from adult liver graft recipients. Children suffer from growth impairment (9) and steroids may influence hepatic regeneration and development of immunologic tolerance (10). The use of CNI bears the risk of developing renal dysfunction after liver transplantation because of its nephrotoxicity (1-3).

IL-2RA specifically bind and block the IL-2R α-chain (which corresponds to CD25), which is present only on the surface of activated T-lymphocytes (11). The IL-2 signal is essential for the activation of lymphocytes; it induces second messenger signals to stimulate T cells to enter the cell cycle and proliferate, resulting in clonal expansion and differentiation. The commercially available IL-2RA are both monoclonal anti-CD25 immunoglobulin G (IgG) antibodies, but their structure and synthesis are different. Daclizumab is a humanized antibody built by total gene synthesis using oligonucleotides (12), whereas basiliximab is a chimeric murine-human antibody (13). The competitive block of IL-2R, and thereby of IL-2-mediated activation, lasts for 4 to 12 weeks, depending on the antibody and the administration protocol (11). The following side effects have commonly been observed in conjunction with the use of IL-2RA: CMV or EBV infection / reactivation, lymphoproliferative disorders, anaphylaxis, fever, opportunist infection, hypotension / hypertension, digestive disturbances, hyperglycemia, hirsutism, pruritus and antibody formation (14).

In the present study, we conduct an analysis of published controlled trials examining the effect of IL-2RA in children after liver transplantation. We would like to investigate whether the use of IL-2RA in addition to concomitant immunosuppressive therapy reduces acute rejection and steroid-resistant rejection after pediatric liver transplantation. We expect that the potential reduction of concomitant medication such as CNI or steroids through the additional therapy with IL-2RA will reduce the long-term adverse drug reaction (ADR) such as kidney failure, disturbance of growth, diabetes mellitus and other metabolic disorders.

**Material and methods**

The methods of literature search, the inclusion and exclusion criteria, outcome measures and methods of statistical analysis were established according to the recommendations in the Cochrane Handbook for Systematic Reviews of Interventions (15, Part 2). We also used the Preferred Items for Systematic Reviews and Meta-Analysis (PRISMA) (16) to structure this report. The methods of this meta-analysis were similar to those used in (17).

*Literature search*
A systematic literature search was performed without language restrictions in December 2012 in the following databases: PubMed, Transplant Library and Cochrane Library. The following search terms were used: "liver transplantation," "interleukin 2 receptor inhibitor/antagonist," "basiliximab," "daclizumab," "zenapax," "simulect," "pediatric," "child," "children," and abbreviations thereof. The keywords were combined with Boolean operators.

*Inclusion and exclusion criteria*
All prospective, controlled pediatric studies and pediatric studies with prospective experimental group and historical control group in which IL-2RA induction therapy in liver transplant recipients was compared with placebo or no add-on were included. A first nonsystematic review of the literature showed that in pediatric liver transplantation IL-2RA are used in addition to standard immunosuppression therapy concepts to reduce other immunosuppressive drugs, such as CNI (3) and corticosteroids (9-10). We have therefore structured this meta-analysis into three separate comparisons as follows: In the first group IL-2RA is added to the experimental group and is compared to no add-on or placebo, while both study arms got equal concomitant immunosuppressive medication. This group is referred to as the *IL-2RA only* comparison in the following. In the second group IL-2RA is additionally combined with delayed CNI in the experimental arm (*delayed CNI* comparison). The third group compared IL-2RA with a standard immunosuppressive protocol with reduced or even dropped steroids in the experimental arm (the *no/low steroids* comparison). Other immunsuppressive medication had to be the same in both treatment arms, e.g., mycophenolate mofetil (MMF).
All retrospective, noncontrolled pediatric studies and pediatric studies with multiorgan transplantation or retransplantation were excluded. Pharmacological studies that did not provide data on clinical outcome measures were excluded as well because of their very short follow-up time. The literature search strategy was designed and performed by two reviewers (N.D.C., A.D.G.). Publications were screened independently by two reviewers (N.D.C., A.D.G.). Disagreement and any discrepancies were resolved by discussion of all four reviewers.

*Outcome measures*
The primary outcomes analysed were acute rejection, steroid-resistant rejection, graft loss and patient death. Secondary outcomes were ADR namely renal dysfunction by need of dialysis or oliguria, de novo malignancy (excluding recurrence of hepatocellular carcinoma), post transplant lymphoproliferative disease (PTLD), infection complications, including fungal, viral and bacterial infections, new onset of metabolic and cardiovascular disorders such as arterial hypertension (HTN), hyperlipoproteinemia (HLP) and posttransplant diabetes mellitus (PTDM).

*Study quality*
The quality items assessed were blinding, randomization, allocation concealment, intention-to-treat (ITT) analysis, completeness of follow-up, and the method of handling missing values. Assessment was performed according to definitions stated in the Cochrane Handbook (15, Ch. 8). Quality of studies was assessed independently by two reviewers (N.D.C., A.D.G.) without blinding to journal and authorship. Furthermore, completeness of follow-up was defined as the number of patients that were not lost to follow-up. We reported completeness of follow-up as stated by the authors. Methods of handling missing values are stated as reported by the authors of the respective study.

*Data extraction*
All available data for the described outcome measures were extracted at all available timepoints from individual trials. In contrast to kidney transplants, it has been shown that morphological signs of rejection in protocol biopsies of transplanted livers without clinical correlates require no treatment and have no long-term ADR (18). Therefore, we only included treated acute rejections in the primary analysis, when the reported acute rejection was stratified into ''treated'' and ''nontreated.'' When data on outcome measures were not provided or studies seemed to be duplicates, the authors were contacted to provide more data. Data extraction was performed by one reviewer (N.D.C.) using a standardized form and checked by two reviewers (A.D.G, C.R.).

*Data analysis*
We expressed the results of dichotomous outcomes as relative risk with values smaller than 1 favoring IL-2RA. When no event was observed in both arms, we excluded it from the corresponding comparison (15, Ch.16.9.3). We performed the analysis using a random effects model, as in case of doubt it makes more sense to use the more general approach (including the fixed-effects model as a special case), which will usually lead to more conservative results (19). For the random effects models the amount of residual heterogeneity (i.e., $\tau^2$) was estimated by the restricted maximum likelihood (REML) method (20). Confidence intervals for $\tau^2$ were obtained by the Q-profile method (21). The model parameters were estimated by way of weighted least squares, with weights equal to the inverse sum of the variance of the estimate and the estimate of the residual heterogeneity. Then Wald-type tests and confidence intervals were obtained for the parameter estimates (20). We analyzed heterogeneity among studies using Cochrane's Q test and calculating $I^2$ to measure the proportion of total variation due to heterogeneity beyond chance (22). We performed subgroup analyses for primary outcomes which had significant results. Subgroups and factors defined a priori were methodological quality of trial (i.e., randomized versus nonrandomized), comparison group, type of IL-2RA, type of CNI, and use of mycophenolate mofetil (MMF). For the primary analysis we pooled effect measurements from trials with different follow-up time; but timepoint of measurement (grouped by 6 months versus 12 months and later) was evaluated in a subgroup analysis. In some of the subgroups a valid analysis was not possible. In order to examine the influence of covariates affecting the direction and/or strength of the relation between dependent and independent variables we used the moderator test. For statistically significant results we calculated the number needed to treat (NNT) describing how many patients are needed to be treated with an intervention, here IL-2RA, to prevent one patient from having one additional bad outcome, here for example acute rejection. Publication bias was assessed using funnel plots (23) and tests for funnel plot asymmetry (20). The R environment for statistical computing (v. 2.11.0) (24) with packages ''meta'' (v. 2.5-0) (25), ''metafor'' (v. 1.4-0) (20), and ''lme4'' (v. 0.999375-37) (26) were used for all analyses.

**Results**

*Literature search*
Database searches yielded 325 entries (see Fig.1), of which 15 were excluded as duplicates. Of the remaining 310 publications that qualified for abstract review, 252 were excluded primarily because they were not controlled trials, the effect of IL-2RA was not investigated, they were not dealing with pediatric patients or they were not conducted in patients undergoing first liver transplantation. Also retrospective studies were excluded. The remaining 58 publications underwent full article review and 38 further publications were excluded. Most common reasons were retrospective studies, other comparator than IL-2RA, studies with adult patients, non controlled studies and reviews. A total of 20 trials qualified for inclusion in this review. 13 studies were excluded because of being duplicates, preliminary reports and follow-up reports of the included studies. One study was excluded because of having no reported follow-up time and the authors did not respond to our requests of further information. Six studies were eventually included for analysis (1-3, 9-10, 27). All trials were obtained as full-text puplications. In case of multiple reports on the same study we cited the most recent full-text publication as the index publication. Two authors of reports were contacted in order to resolve ambiguities. One author answered, the other author did not respond.

*Included studies*
Table 1 shows the characteristics of the included studies. Three trials (1-2, 27) compared IL-2RA to no treatment without modification of concomitant immunosuppressive medication (*IL-2RA only* comparison). Only one trial (3) compared IL-2RA in combination with delayed CNI to no treatment with standard immunosuppression (*delayed CNI* comparison). Two trials (9-10) compared IL-2RA in combination with early discontinuation or reduction of steroids to no treatment with standard immunosuppression (*no/low steroids* comparison). One study (9) excluded patients with a severe renal dysfunction and another trial (10) excluded children with autoimmune hepatitis. In two-thirds of the included studies basiliximab was used for induction (1, 9-10, 27) and another two-thirds of the trials used tacrolimus as CNI (2-3, 9-10). Two of the six studies used MMF (2-3) and all but the experimental arms of the studies of *no/low steroids* comparison (9-10) used steroids as concomitant medication. Most trials had a study duration of 12 months or more (3, 9-10, 27).

*Quality of included studies*
Table 2 shows the quality assessment of the included studies. None of the studies were classified as blinded, whereas two of them did not report the status of blinding and were classified as „not stated". Two studies (3, 9) were prospective and randomized, a third study (2) was entirely prospective and three studies (1, 10, 27) had a prospective experimental group and historical control group. Of the randomized trials, allocation concealment was found to be adequate in none of the studies, but unclear in one study (9) and inadequate (3) in another. For the four nonrandomized studies (1-2, 10, 27) allocation concealment was not applicable. ITT analysis was stated and performed in one study (9) and was assessed as adequate. In three further trials (2, 10, 27) ITT analysis was assumed and considered adequate because the authors reported on all patients at the endpoints of the study. One study (1) reported ITT analysis but it was assessed as inadequate. According to the definition given in (15, Ch. 16.2.1), the authors of that trial did a per-protocol (PP) analysis. In only one trial (3) we could not assess ITT analysis and it was therefore classified as „not stated". None of the authors stated how missing values were handled. Only two studies (9-10) described completeness of follow-up.

*Application of IL-2RA*
Table 3 summarizes the immunosuppressive therapy of included pediatric trials. Basiliximab was used in four studies (1, 9-10, 27). All of them administered basiliximab on postoperative day-0 and 4 in a dosage of 10mg for children with a weight below 30 kg or a dosage of 20mg for children with a weight of more than 30 kg. In addition Spada et al. 2006 (9) administered basiliximab on postoperative days-8 until -10 if the recipient had lost more than 70 ml/kg fluid from the abdominal drains, because basiliximab crosses into the ascitis fluid (3). Daclizumab was used as induction therapy adapted to patient's weight (1mg/kg) in two trials (2-3) and was therefore administered on postoperative day-0. Schuller et al. 2005 (2) in addition gave a second dose on postoperative day-14. In the immunosuppressive concepts of the studies discussed here, the authors tried to limit the overall immunosuppression (3). Daclizumab has a half-life of 99hr and it's loss in ascitic fluid has been only weakly correlated to the monoclonal antibody clearance (3, p 2040). Heffron et al. 2003 (3) used it in the first week after transplantation instead of CNI, whereas Schuller et al. 2005 (2) used daclizumab induction to reduce the target level of tacrolimus from the beginning.

*Definition of primary outcomes*
Most studies defined acute rejection as a rejection episode (1-2, 9-10), confirmed by liver biopsy (1, 3, 9-10, 27), and for which therapy was given (2-3, 9-10, 27). Some trials described a clinical and laboratory diagnosis in addition (3, 27). The severity of AR was graded by using the Banff criteria (28) in two studies (9, 27). A steroid-resistant acute rejection was defined as not past by using steroids (2-3, 9, 27) and therefore a treatment with for example OKT3 (3, 9) was needed. Some studies proved it by biopsy (3, 27). Spada et al. 2006 (9) also used CNI in standard dose first before adding steroids.

*Follow-up time of included studies*
Follow-up times varied from 6 up to 52 months. They also differed between control and experimental group and were not necessarily identical for all outcomes. Because of the different follow-up times a comparison is difficult, but the first 6 months are the crucial period in which IL-2RA is acting. The long time effects on outcome of patients should be measured over years. We have not found enough data on long-term outcomes. Most follow-up is about 3 years only.

*Primary ouctomes*
*Acute rejection.* Reduction of acute rejection favored the use of IL-2RA (RR 0.38; CI 0.22–0.66; p = 0.0005; 6 trials; Fig. 2). The effect is also seen in the subgroup of randomized trials (RR: 0.31; CI 0.20–0.47; p < 0.0001; 3 trials), but is not statistically significant in nonrandomized studies (RR: 0.46; CI 0.18–1.18; p = 0.1039; 3 trials). The relative risk (RR) of all studies had a statistically significant heterogeneity (p=0.0126) which is due to the study of Gibelli et al. 2004 (1). Omitting the study by Gibelli et al. 2004 (1) the risk reduction is larger (RR 0.30; CI 0.21–0.43; p < 0.0001; 5 trials). Considering the three pre-specified subgroups of studies, we have data on the *IL-2RA only* comparison (RR: 0.44; CI 0.19–1.002; p = 0.0507; 3 trials) and on the *no/low steroids* comparison (RR: 0.31; CI 0.12–0.84; p = 0.0211; 2 trials). Only the *no/low steroids* comparison was significant in reduction of acute rejection, favoring the use of IL-2RA. Stratifying studies by follow-up time showed a statistically significant reduction of acute rejections with IL-2RA at 12 months and later (RR 0.31; CI 0.21–0.45; p < 0.0001; 4 trials), but not at 6 months (RR 0.51; CI 0.16–1.66; p = 0.2654; 2 trials). Furthermore, subgroup analysis stratified by the type of IL-2RA used showed a statistically significant effect of both basiliximab (RR 0.44; CI 0.21–0.92; p = 0.0299; 4 trials) and daclizumab (RR 0.29; CI 0.18–0.47; p < 0.0001; 2 trials). The subgroup with daclizumab induction therapy got additional MMF as immunosuppressive concomitant medication, it showed a lower p-value and showed no statistically significant heterogeneity. Stratifying trials by type of CNI used showed a statistically significant effect of tacrolimus (RR 0.30; CI 0.19–0.46; p < 0.0001; 4 trials) but not for cyclosporine A (RR 0.53; CI 0.20–1.40; p = 0.1999; 2 trials). Finally we analysed the subgroup stratified by control group. Studies with retrospective control group showed no statistically significant reduction of acute rejection (RR 0.46; CI 0.18–1.18; p = 0.1039; 3 trials) in comparison to these with prospective control group (RR 0.31; CI 0.20–0.47; p < 0.0001; 3 trials). The number needed to treat (NNT) is 3.6, which means that four children after liver transplantation have to be treated with IL-2RA in addition to standard immunosuppressive therapy to prevent one patient of having an acute rejection. Four studies defined acute rejection clinically and confirmed it by biopsy (1-3, 27). Analysis of this subgroup showed a statistically significant reduction of acute rejection (RR 0.40; CI 0.21–0.76; p = 0.0052; 4 trials). One trial (9) used the term of an episode of acute rejection as outcome measurement and another study (10) did not state a definition. None of the trials were taking protocol biopsies.

*Steroid-resistant rejection.* All trials reported data on steroid-resistant rejection. One study (9) reported about no steroid-resistant rejection in both arms, so that we excluded it from analysis. IL-2RA in addition to standard double or triple immunsuppressive therapy after liver transplantation in children did not reduce steroid-resistant rejection statistically significantly (RR 0.44; CI 0.19–1.03; p = 0.0594; 5 trials). If we exclude one of the older studies (1) with the most extreme effect from the analysis, we get a statistically significant reduction of steroid-resistant rejection without significant heterogeneity (RR 0.34; CI 0.14–0.79; p = 0.0123; 4 trials). Stratifying trials by randomization status (randomized subgroup: RR 0.18; CI 0.04–0.74; p = 0.0177; 2 studies (2-3) and nonrandomized subgroup: RR 0.65; CI 0.24–1.78; p = 0.3971; 4 trials (1, 9-10, 27)) and comparison did not show statistically significant effects (*IL-2RA only* comparison: RR 0.77; CI 0.30–1.98; p = 0.5894; 3 trials (1-2, 27); *delayed CNI* comparison (3) and *low/no steroids* comparison (10) only one study each) except the subgroup of randomized studies. However, we saw a statistically significant reduction of steroid-resistant rejection in studies (3, 10, 27) with follow-up measurements at 12 months and later (RR 0.33; CI 0.12–0.89; p = 0.0281; 3 trials; Fig.3), but not for 6 months (RR 0.93; CI 0.18 –4.67; p = 0.9269; 2 trials (1-2)). There was a statistically significant reduction in steroid-resistant rejection by using IL-2RA in subgroups using tacrolimus (RR 0.17; CI 0.05–0.57; p = 0.0041; 3 trials (2-3, 10)) and daclizumab induction therapy combined with additional MMF dose and prospective control group (RR 0.18; CI 0.04–0.74; p = 0.0177; 2 trials (2-3)).

*Graft loss and patient death.* Four studies (3, 9-10, 27) reported data on graft loss and patient death. Neither graft loss (RR 0.65; CI 0.34–1.21; p = 0.1737; 4 trials) nor patient death (RR 0.61; CI 0.27–1.37; p = 0.2296; 4 trials) were statistically significantly reduced by using additional IL-2RA in combination to standard immunosuppressive medication in the observation period. In the forest plot one study (9) is prominently deviating from the remaining studies. After excluding this study from the analysis we saw a statistically significant result for reducing graft loss by the use of IL-2RA (RR 0.44; CI 0.21–0.92; p = 0.0298; 3 trials), but not for death (RR 0.42; CI 0.16–1.12; p = 0.0831; 3 trials). However, all analyses show a trend towards a lower incidence of graft loss and patient death in the experimental group using IL-2RA in addition to standard double-drug or triple-drug therapy.

*Secondary outcomes: Side effects and subgroups*
It was not possible to collect enough data to analyse secondary outcomes, namely de novo malignancy, PTDM or HLP. Four studies reported on renal dysfunction (1, 2, 9, 27); data analysis of this outcome showed a slight reductive tendency but no statistically significant reduction by using IL-2RA (RR 0.96; CI 0.60–1.54; p = 0.8683; 4 trials). Furthermore three studies (1, 9, 27) reported on new onset arterial hypertension (HTN). Analysis showed no significant reduction of HTN by using IL-2RA but a reductive tendency (RR 0.85; CI 0.60–1.21; p = 0.3731; 3 trials). Three studies (2, 9, 27) reported on PTLD but one of them (2) yielded no event in both arms, so that we excluded it from analysis. PTLD was not reduced by using additional IL-2RA therapy (RR 1.6; CI 0.20–12.67; p = 0.6587; 2 trials), on the contrary, it showed a higher RR in the experimental group. Two studies (9, 27) reported on infection complications and outcomes were also reported on subgroups named viral, bacterial and fungal infections. Additional IL-2RA therapy with standard immunosuppressive medication did not reduce infection complications statistically significantly (infection complications: RR 0.80; CI 0.60–1.07; p = 0.1363; 2 trials; viral infection: RR 1.06; CI 0.62–1.80; p = 0.8356; 2 trials; fungal infection: RR 1.15; CI 0.46–2.87; p = 0.7624; 2 trials; bacterial infection: RR 0.68; CI 0.34–1.37; p = 0.2838; 2 trials). Infection complications and bacterial infection showed a reductive tendency but viral and fungal infection were more frequent in the control group. Due to limited data we were unable to do subgroup analyses except for the *IL-2RA only* comparison. The subgroup analysis of AE of the *IL-2RA only* comparison showed no statistically significant reduction by using IL2-RA in any of the secondary outcomes renal dysfunction, HTN or PTLD.

**Discussion**

The use of IL-2RA in addition to standard double-drug or triple-drug therapy significantly lower the risk of acute rejection in pediatric patients after liver transplantation. Acute rejection rate is reduced by two thirds through the use of IL-2RA (RR 0.38). These results are similar to those we found in our meta-analysis in adult liver transplant recipients (17). The relative risk of all studies has a significant heterogeneity which is introduced by the study of Gibelli et al. 2004 (1). Most of the subgroup analyses support a statistically significant reduction of acute rejection through the additional use of IL-2RA, all subgroup analyses showed a substantial reduction by at least 50%.
The use of IL-2RA in addition to standard double-drug or triple-drug therapy also shows a substantial reduction of steroid-resistant rejection after pediatric liver transplantation (RR 0.44). If we exclude one of the older and prominently deviating studies (1) from analysis, we get a statistically significant reduction of steroid-resistant rejection without significant heterogeneity. Subgroup analysis stratified by measurement time at 12 months and later, randomized subgroup, as well as a subgroup of only prospective controlled trials observed significant reduction of steroid-resistant rejection through the use of IL-2RA.
Although the risk of acute rejection is substantially reduced when IL2-RA is applied, we did not observe a statistically significant reduction in graft loss or patient death. Observed trends suggested that the number of patients may be too small to observe significant effects, but we see a clinically relevant reduction of patient death (RR 0.61) and graft loss (RR 0.65) by about one third. These results are similar to those we found in adult liver transplantated patients (17).
We also looked at the possibility of reducing concomitant immunosuppressive medication when using IL-2RA because most published studies explored this effect. We could classify published studies into three different experimental immunosuppressive regimes, namely the *IL-2RA only* comparison (1-2, 27), the *delayed CNI* comparison (3), and the *no/low steroids* comparison (9-10). Stratifying trials by comparison, there is only the *IL-2RA only* comparison and the *no/low steroids* comparison to analyse. The *no/low steroids* comparison shows a statistically significant reduction of acute rejection favoring the use of IL-2RA. In the analysis of other primary outcomes the number of studies in each comparison is too small, so that we find no statistically significant effect in any. We see a clinically relevant reductive effect in the *IL-2RA only* and *no/low steroids* comparisons of the risk of steroid-resistant rejection, patient death and graft loss through the additional use of IL-2RA in the experimental group.
Compared to other types of pediatric solid organ transplantations, we see a similar reductive tendency of AR and SRAR in pediatric renal recipients receiving induction therapy with IL-2RA. Swiatecka-Urban et al. 2001 (29), a study with retrospective control group, compared basiliximab induction and tacrolimus with no treatment in pediatric renal recipients. The use of basiliximab induction in addition to tacrolimus and steroids reduce the risk of rate of AR (basiliximab group (BG): 26% vs. non-BG: 43%; p = 0.36) and rate of SRAR (BG: 8.7% vs. non-BG: 12.5%; p = 0.68). No patient deaths were observed within one year follow-up time. The one-year graft survival rate was higher in the induction group (BG: 87.5% vs. 75% in non-BG; p = 0.45). These results are comparable with results of other studies, for example Vester et al. 2001 (30). In this prospective study using basiliximab as induction combined with cyclosporin A and prednisone one year patient survival rate was 100%, graft survival rate was 95%, AR episodes were observed in six patients and two SRAR were observed. Also a historical controlled study comparing basiliximab with no medication and triple baseline immunosuppression with cyclosporine or tacrolimus, prednisone, MMF showed a reduction of AR to 10% in the induction group compared to 38% in the control group (31). There are very few studies using IL-2RA in pediatric patients after lung or heart transplantation. We found occasional controlled studies while most publications were reviews. One controlled study reported about a six-months AR incidence of 30% in the daclizumab group vs. 60% in the control group (32). IL-2RA are also used in pediatric patients after heart transplantation (33-34). It seems to reduce acute rejection if basiliximab is given before transplantation (35) and reduced AR in critically ill children with heart transplantation (36). IL-2RA induction therapy after lung and heart transplantation showed a reductive tendency of AR along with an acceptable safety profile, but the reductive tendency in pediatric patients seems to be stronger after liver transplantation compared to published data on renal, heart or lung transplant recipients.

The following side effects were observed after basiliximab application in about 20% of pediatric patients by the EPAR: urinary tract infections, hypertrichosis, rhinitis, fever, hypertension, upper respiratory tract infection, viral infection, sepsis and constipation (37, p. 2). The EPAR reported about side effects of daclizumab such as insomnia, tremor, headache, hypertension, dyspnoea, constipation, diarrhoea, vomiting, nausea, dyspepsia, musculoskeletal pain, oedema, impaired healing and post-traumatic pain being observed in more than one out of ten patients (38, p. 2). Of these named potential side effects, the analysed studies reported mostly about metabolic disorders, for example HTN, and observed infection complications which were not reported in detail. Major side effects as lymphoma were observed rarely.

Due to the limited amount of data we were unable to perform subgroup analyses except for the *IL-2RA only* comparison. Also it was not possible to collect enough data to analyse secondary outcomes, namely de novo malignancy, PTDM or HLP. Analysis of included studies shows a reductive tendency of renal dysfunction, new onset posttransplant arterial hypertension and infection complication especially bacterial infections in experiemtal group which uses IL-2RA in addition to standard immunsuppressive therapy. The subgroup analysis of the *IL-2RA only* comparison showed no statistically significant reduction by using IL2-RA in any of the secondary outcomes called renal dysfunction, HTN or PTLD.

In the published European Public Assessment Report (EPAR) about simulect (basiliximab), the weight limit for a higher dose is 35 kg (37). Basiliximab was studied in pediatric and adult kidney transplanted patients. It was given on postoperative day-0 and 4 (37). According to the EPAR, daclizumab should be given in 1mg/kg on postoperative day 0, 14, 28, 42, and 56 after kidney transplantation (38). By comparing the mode of application between analysed studies and EPAR statement we see that the authors have used the common dose of basiliximab. Only Spada et al. 2006 (9) added a third dose, which might hold the level of IL-2RA intracorporeal. The limit of weight for a higher dose was lower than proposed by EPAR. Daclizumab was given only once or twice after liver transplantation. Giving a drug in one or two single doses is a good practical application, promotes the compliance and may shorten the day of hospitalization.

**Strengths and limitations**

The main limitation of this review is the small number of randomized controlled trials, even compared to trials in kidney transplantation (39), and our systematic review and meta-analysis of adult patients after liver transplantation (17). The low number of studies makes it difficult to acquire enough data to demonstrate statistical significance. Corresponding to our experience with studies of adult liver transplant recipients (17) we decided to include not only randomized trials but also nonrandomized controlled trials and studies with prospective experimental group and retrospective control group in this review. Half of them compared IL-2RA to no add-on. The other half explored the effects of reduced or delayed concomitant immunosuppression. Therefore, we decided to include those studies in order to increase the total number of included trials. We also allocated them to predefined comparisons of concomitant medication. Furthermore, we included and pooled studies that used a different type of IL-2RA, had different concomitant medication (type of CNI and MMF), or had different follow-up times. Because all these differences may be sources of heterogeneity, it was planned to explore differences of effect by performing subgroup analyses. Because of the small number of included studies some studies dominate the results as we have seen in analyses including Gibelli et al. 2004 (1) or Spada et al. 2006 (9). Both studies met the inclusion criteria. Due to the paucity of data on secondary outcomes we were only able to extensively analyze the primary endpoints. Another problem was the insufficient detailed reporting of outcomes; this was noticed most evidently regarding the adverse drug reaction of immunosuppression. Few studies give data on complications and ADR, but also these were measured or grouped differently in the various trials. We endeavored to overcome this limitation by grouping data on side effects into broader categories, but this may further limit the interpretation of the results. Although we attempted to minimize publication bias by searching for and including data from different databases. Nonetheless, this systematic review and meta-analysis gives us a first impression of the evidence and the order of magnitude of the effect of using IL-2RA as an induction therapy in addition to standard double-drug or triple-drug therapy in pediatric liver transplant recipients. For further analysis we require more studies, but we do not expect more data to accumulate over the next years. In order to gain information on long-term effects of reduced or delayed concomitant immunosuppression, which is urgently needed in pediatric liver transplant recipients, more prospective controlled trials are needed.

**Clinical implications**

Four pediatric patients would need to be treated with IL-2RA to prevent one acute rejection (NNT ≈ 4). The risk reduction for acute rejection is higher than would be expected from experience with adult liver transplantation (17) which could be a result of differences in the pediatric metabolism. We have no evidence for a difference in effectiveness between basiliximab and daclizumab in reducing the risk of rejection. In conclusion, the use of IL-2RA reduces the risk of acute rejection without a significant increase of harmful effects. This effect may allow for reduction of coimmunosuppression to avoid the adverse drug reaction of CNI or steroids. Also we observe a substantial reduction of the risk of steroid-resistant rejection, patient death and graft loss by using IL-2RA in addition to standard double-drug or triple-drug therapy, and therefore we should value this result as clinical relevant.

**Acknowledgement**

**Table 1. Characteristics of included trials stratified by the three pre-specified comparison groups**

| Trial (Author & Year) | Patient Subgroup² | Sample Size | | Age^ | | Sex (male) | | Typ of IL-2RA | Control Substance | CNI | MMF | Follow-up§ |
|---|---|---|---|---|---|---|---|---|---|---|---|---|
| | | Exp | Cont | Exp | Cont | Exp | Cont | | | | | |
| *IL-2RA only* comparison: IL-2RA vs placebo/no treatment | | | | | | | | | | | | |
| Ganschow 2005 | | 54 | 54 | 4.2 (0.3-8.9)" | matched | ns | ns | Bas | no | Cya | no | 36 |
| Gibelli 2004 | | 28 | 28 | 3 (1.3-16)" | matched | ns | ns | Bas | no | Cya | no | 6 |
| Schuller 2005 | | 18 | 12 | 3.95 ± 0.33 | 3.9 ± 0.26 | 9 | 4 | Dac | no | Tac | yes | 6 |
| *delayed CNI* comparison: IL-2RA and delayed and/or reduced CNI vs placebo/no treatment and standard immunosuppressive co-medication | | | | | | | | | | | | |
| Heffron 2003 | | 61 | 20 | 6.8 ± 6.3 | 5.3 ± 6.6 | 24 | 7 | Dac | ns | Tac | yes | 24 |
| *no/low steroids* comparison: IL-2RA and minimized steroids or no steroids vs placebo/no treatment and standard immunosuppressive co-medication | | | | | | | | | | | | |
| Spada 2006 | renal function | 36 | 36 | 2.9 (1.5-4.3)" | 2.8 (1.5-4.3) | 18 | 15 | Bas | no | Tac | no | 12 |
| Gras 2008 | no auto-immun hepatitis | 50 | 34 | 1.7 (0.4-14.0)" | 2.0 (0.4-14.0) | 27 | 16 | Bas | no | Tac | no | 36 |

**Abbreviations:** Exp – experimental group; Cont – control group; IL-2RA– interleukin-2 receptor antagonist; CNI – calcineurin inhibitor; MMF – mycophenolat mofetil; Bas – basiliximab; Dac – daclizumab; Cya – cyclosporine A; Tac – tacrolimus; ns – not stated; vs – versus.
^ - Age is given in mean ± standard deviation if available. " - Age is given in mean with (minimum – maximum).
§ - Length of follow-up, time is given in months. ²- Patient Subgroup: Special inclusion criteria used by the authors.

**Table 2. Summary of quality assessment of included trials**

| Trial (Author & Year) | Blinding | Randomized | Control Group | Allocation Concealment | ITT Analysis | Missing Values | Completeness of follow-up § Exp (%) | Cont (%) | Month |
|---|---|---|---|---|---|---|---|---|---|
| *IL-2RA only* comparison: IL-2RA vs placebo/no treatment | | | | | | | | | |
| Ganschow 2005 | no | no | historical | na | yes^ (ns) | ns | ns | ns | 28-52 |
| Gibelli 2004 | no | no | historical | na | yes* (PP) | ns | ns | ns | 6 |
| Schuller 2005 | no | no | concurrent+ | na | yes^ (ns) | ns | ns | ns | 6 |
| *delayed CNI* comparison: IL-2RA and delayed and/or reduced CNI vs placebo/no treatment and standard immunosuppressive co-medication | | | | | | | | | |
| Heffron 2003 | ns | yes | concurrent | inadequate | ns | ns | ns | ns | 24 |
| *no/low steroids* comparison: IL-2RA and minimized steroids or no steroids vs placebo/no treatment and standard immunosuppressive co-medication | | | | | | | | | |
| Spada 2006 | ns | yes | concurrent | unclear | yes | ns | 90 | 90 | 12 |
| Gras 2008 | no | no | historical | na | yes^ (ns) | ns | 100 | 100 | 36 |

**Abbreviations:** ITT – intention-to-treat;, na – not applicable, ns – not stated; PP – per-protocol; Exp – experimental group; Cont – control group; vs - versus.
^ - ITT-analysis is assumed, because the author reported about at least one analysis with the total number of included patients.
* - author reported that ITT analysis was performed, but also stated conditions that must be met for patient to be included in analysis, such as "patient received at least one dose of medication" and/or "at least one follow-up available". § - as stated by authors or calculated from available data. + - prospective study.

**Table 3. Immunosuppressive therapy of included pediatric trials**

| Trial (Author & Year) | IL-2RA Type & Dosage | CNI Type, First Day of Therapy, Dosage & Target Level (tl) | | Corticosteroids Dosage† | | MMF Dosage |
|---|---|---|---|---|---|---|
| | | Exp | Cont | Exp | Cont | |
| *IL-2RA only* comparison: IL-2RA vs placebo/no treatment | | | | | | |
| Ganschow 2005 | Basiliximab POD 0&4 i.v. 10mg (KG < 30kg) 20mg (KG > 30kg) | Cyclosporine A tl 150 - 200 μg/L after one year tl 80 - 100 μg/L | | Prednisolone 60 mg/m² after one week 30 mg/m² thereafter weekly reduction about 5 mg/m² break off after one year | | no |
| Gibelli 2004 | Basiliximab POD 0&4 i.v. 10mg (KG < 30kg) 20mg (KG > 30kg) | Cyclosporine A 7 – 13mg/kg/d tl 850 – 1000 mg/dL | Cyclosporine A 5 – 7 mg/kg/d tl 850 – 1000 mg/dL | steroids | | no |
| Schuller 2005 | Daclizumab POD 0 & 14 i.v. 1mg/kgKG | Tacrolimus 0.2 mg/kg/d tl 10-12 ng/mL Start: POD 3 | Tacrolimus 0.2 mg/kg/d tl 6-8ng/mL Start: POD 3 | Methylprednisolone 20mg/kg Start: POD 0 fast reduction break off after 4th months | | 1200mg/m²/d Start: POD 14 |
| *delayed CNI* comparison: IL-2RA and delayed and/or reduced CNI vs placebo/no treatment and standard immunosuppressive co-medication | | | | | | |
| Heffron 2003 | Daclizumab POD 0 i.v. 1mg/kgKG | Tacrolimus 0.15/mg/kg/d tl 10-14ng/mL Start: POD 7 | Tacrolimus 0,15/mg/kg/d tl 10-14ng/mL Start: POD 0 | Methylprednisolon: POD 0: 20mg/kg/d, POD 6: 0,3mg/kg/d | | 30mg/kg/d, p.o. |
| *no/low steroids* comparison: IL-2RA and minimized steroids or no steroids vs placebo/no treatment and standard immunosuppressive co-medication | | | | | | |
| Spada 2006 | Basiliximab POD 0 & 4 & POD 8-10 i.v. 10mg (KG < 35kg) 20mg (KG > 35kg) | Tacrolimus 0.04 mg/kg/d, p.o. tl 1th month 10-15 ng/mL tl 2th-3th month 10-15ng/mL tl 4th-6th month 6-8ng/mL thereafter 5-7ng/mL | | Methyl-prednisolone i.o. 10mg/kg i.v. | Methylprednisolone i.o. 10mg/kg i.v. POD 1-6: 2mg/kg/d POD 7: 1mg/kg/d break off after 3th - 6th months maximum: 40mg | no |
| Gras 2008 | Basiliximab POD 0 & 4 i.v. 10mg (KG < 35kg) 20mg (KG > 35kg) | Tacrolimus 0.2 mg/kg/d, p.o. Start: POD 0 tl 1th month 8-12 ng/mL thereafter 5-8 ng/mL | | no | Methylprednisolone 10mg/kg/i.v. POD1- 6: 2 mg /kg/d i.v. POD7 – 13: 1mg/kg/d p.o. POD14-20: 0.75 mg/kg/d POD 21-28: 0.5 mg/kg/d thereafter 0.25 mg/kg/d POD 90: 0.25mg/kg/d (cave: alternative therapy in 2th -6th months) | 20 mg/kgKG/d (only the first 9 children took it) |

**Abbreviations:** i.o. – intraoperative; POD – postoperative day; Tac – tacrolimus; tl – target level; KG - body weight; kgKG – kilogamme body weight; Exp – experimental group; Cont – control group; na - not applicable (drugs were not used), ns – not stated (drugs were used but not specified), vs – versus.
† - All trials used methylprednisolone intraoperatively. In the postoperative period they used methylprednisolone or prednisolone.

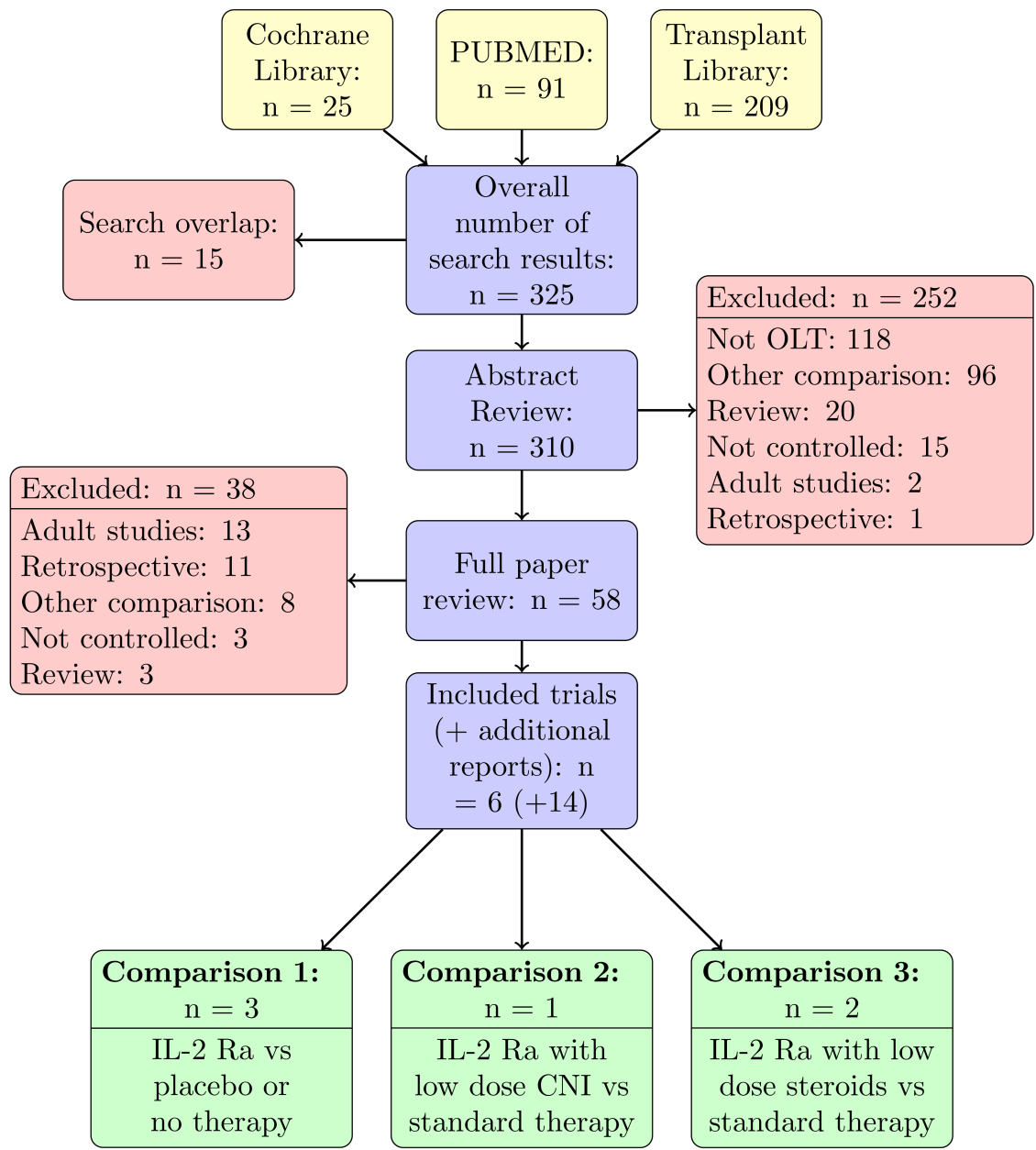

**Fig. 1.** Flow chart of systematic review. IL-2 RA - interleukin-2 receptor antagonist (according to PRISMA (16)) Abbreviations: CNI - calcineurin inhibitor; OLT - orthotopic liver transplantation
(150.8mm x 165.9mm)

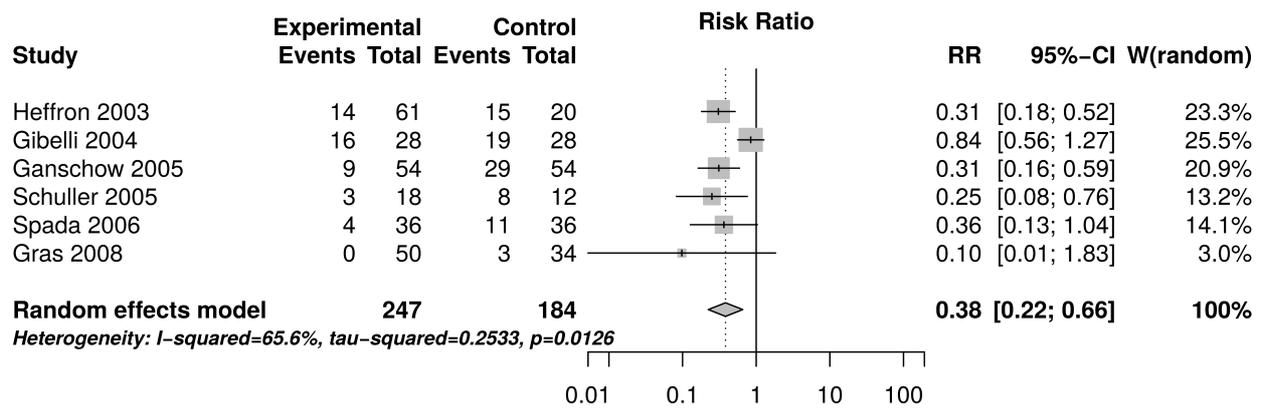

**Fig. 2.** Forest plot of acute rejection of all included studies. The forest plot shows a reduced relative risk of acute rejection for pediatric patients which have used IL-2RA (experimental group). The result is significant, but also shows significant heterogeneity (p = 0.0126). Abbreviations: RR - relative risk; 95%-CI - 95% confidence interval; p – p value for test of heterogeneity; Experimental - experimental group, Control - control group.

(210mm x 73.8mm)

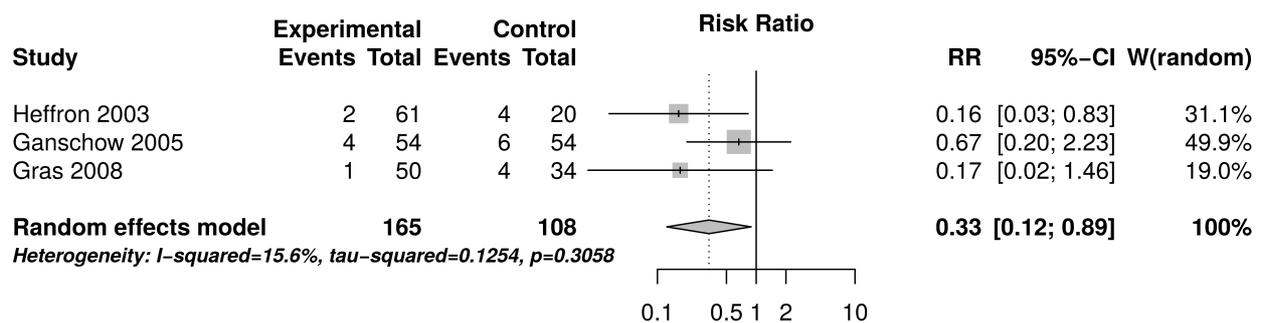

**Fig. 3.** Forest plot of steroid-resistant rejection stratified by follow-up measurement, here follow-up 12 months and later. The forest plot shows a reduced relative risk of steroid-resistant rejection for pediatric patients which have used IL-2RA (experimental group). The result is significant and shows no significant heterogeneity (p = 0.3058). Abbreviations: RR - relative risk; 95%-CI - 95% confidence interval; p – p value for test of heterogeneity; Experimental - experimental group, Control - control group.

(210mm x 65.8mm)

33. REDDY SC, LAUGHLIN K, WEBBER SA. Immunosuppression in Pediatric Heart Transplantation: 2003 and Beyond. Curr Treat Options Cardiovasc Med 2001: 5(5): 417-428. Abstract
34. PIETRA BA, BOUCEK MM. Immunosuppression for pediatric cardiac transplantation in the modern era. Prog Pediatr Cardiol 1. 2000:11(2):115-129. Abstract
35. GRUNDY N, SIMMONDS J, DAWKINS H, REES P, AURORA P, BURCH M. Pre-implantation basiliximab reduces incidence of early acute rejection in pediatric heart transplantation. J Heart Lung Transplant. 2009: 28(12): 1279-1284. doi: 10.1016/j.healun.2009.09.001. Epub 2009 Oct 28. Abstract
36. FORD KA, CALE CM, REES PG, ELLIOTT MJ, BURCH M. Initial data on basiliximab in critically ill children undergoing heart transplantation. J Heart Lung Transplant. 2005: 24(9): 1284-1288. Abstract
37. EUROPEAN MEDICINES AGENCY (EMEA): European Public Assessment Report (APAR) Simulect - EPAR summary for the public. 2009. URL: http://www.ema.europa.eu/docs/en_GB/document_library/EPAR_-_Summary_for_the_public/human/000207/WC500053541.pdf (accessed Sep. 4, 2014).
38. EUROPEAN MEDICINES AGENCY (EMEA): European Public Assessment Report (APAR) Zenapax - EPAR summary for the public. 2008. URL: http://www.ema.europa.eu/docs/en_GB/document_library/EPAR_-_Summary_for_the_public/human/000198/WC500057570.pdf (accessed Sep. 4, 2014).
39. WEBSTER AC, PLAYFORD EG, HIGGINS G, CHAPMAN JR, CRAIG JC. Interleukin 2 receptor antagonists for renal transplant recipients: a metaanalysis of randomized trials. Transplantation 2004: 77: 166-176.
18